# In-situ liquid SAXS studies on the early stage of calcium carbonate formation

*Ahmed S. A. Mohammed[1,2,3,‡], Agnese Carino[4,‡], Andrea Testino[4,*], Mohammad Reza Andalibi[4,5], Antonio Cervellino[1].*

[1]Paul Scherrer Institute (PSI), Swiss Light Source (SLS), CH-5232 Villigen, Switzerland

[2]Physik-Institut der Universität Zürich, Winterthurerstrasse 190, CH-8057 Zurich, Switzerland

[3]Physics Department, Faculty of Science, Fayoum University, 63514 Fayoum, Egypt.

[4] Paul Scherrer Institute (PSI), ENE, CH-5232 Villigen, Switzerland

[5] École Polytechnique Fédérale de Lausanne (EPFL), STI, CH-1015 Lausanne, Switzerland

[‡]Both authors contributed equally to this work.

* To whom correspondence should be addressed: andrea.testino@psi.ch

ABSTRACT

Calcium carbonate is a model system to investigate the mechanism of solid formation by precipitation from solutions, and it is often considered in the debated classical and non-classical nucleation mechanism. Despite the great scientific relevance of calcium carbonate in different areas of science, little is known about the early stage of its formation. We therefore designed contactless devices capable to provide informative investigations on the early stages of the precipitation pathway of calcium carbonate in supersaturated solutions using classical scattering methods such as Wide-Angle X-ray Scattering (WAXS) and Small-Angle X-ray Scattering (SAXS) techniques. In particular, SAXS was exploited for investigating the size of entities formed from supersaturated solutions before the critical conditions for amorphous calcium carbonate (ACC) nucleation are attained. The saturation level was controlled by mixing four diluted solutions (i.e., NaOH, $CaCl_2$, $NaHCO_3$, $H_2O$) at constant $T$ and $pH$. The scattering data were collected on a liquid jet generated about 75 sec after the mixing point. The data were modeled using parametric statistical models providing insight about the size distribution of denser matter in the liquid jet. Theoretical implications on the early stage of solid formation pathway are inferred.

1. INTRODUCTION

Understanding the early stage in the precipitation pathway is of fundamental relevance in order to achieve an appropriate control at the macroscopic level on the solid formation, for instance in terms of morphology, physicochemical properties, and crystalline phase.[1–6] $CaCO_3$ is a model system, archetype of several sparsely soluble inorganic materials. The precipitation pathway includes the nucleation and growth of a solid phase from supersaturated solutions. At the early stage, even the physical nature of the entities with a density higher than that of the solvent is a matter of debate.[7]

The precipitation is a complex process. Homogeneous primary nucleation is the first elementary process to consider in solid formation pathway.[8] There are two descriptions developed for such phenomenon occurring in solutions: the classical nucleation theory (CNT) and non-classical nucleation theory (NCNT).[5,9–11] These theories have been recently reviewed.[12] In both cases, the formation of embryos or clusters starts with a reacting solution of dissolved aqueous species, triggered by a change of the system saturation. Afterwards, these entities grow and the reaction may follow several pathways over a period of time, according to the Ostwald's rule of stages, until the final thermodynamic stable product is achieved.

In the formulation of the CNT, a series of approximation and simplifications are assumed with the aim to obtain a simplified mathematical framework. One of the most relevant assumptions is the so-called capillary approximation that leads to the definition of a landscape of the Gibbs free energy vs. embryo size characterized by only one maximum, which coordinates identify the activation energy and the critical size for nucleation. Any alternative landscapes identify a NCNT where, for instance, an arbitrary mathematical equation is considered for the surface energy as a function of clusters size. In this case, the Gibbs free energy landscape can admit more than one point where the derivative is zero, e.g. a minimum before the critical size. The existence of a minimum in the Gibbs free energy identifies a privileged size for clusters of subcritical dimensions. In fact, according to the NCNT, in a supersaturated solution, stable and well-defined clusters (called pre-nucleation clusters) exist and they may aggregate or coagulate into larger entities. In the specific case of calcium carbonate, the first solid product has an amorphous nature (amorphous calcium carbonate, ACC), which eventually transforms toward more thermodynamically stable crystalline phases.[13,14] Some authors claim that such intermediate amorphous phase has a kind of local ordering, which depends on the *pH* value and defines the crystal phase that may be formed by its evolution, naming it as proto-calcite or proto-vaterite.[11] According to our recent finding,[15] *pH* has an influence on the calcium-to-carbonate ratio in the amorphous phase as well as in the pre-nucleation entities, which are in dynamic equilibrium with the solution. This ratio might be the reason of the preferential phase obtained upon crystallization.

Indeed, the existence of sub-critical embryo is expected even following the CNT reasoning with a statistical distribution of sizes, which is influenced by the saturation level. Therefore, other authors claim that there is no preferential size and thus the nucleation is classical.[16] The common point is that it is generally accepted that subcritical embryos or clusters exist. Using advanced analytical tools, such as cryoTEM or analytical ultracentrifugation (AUC), small entities were identified.[11,17] Nevertheless, sample manipulation is necessary and considering that such entities are in equilibrium with the liquid in which they are formed, any small physicochemical modification can strongly influence their stabilization/destabilization. As a consequence, the investigations of these labile entities need to be done in-situ, without modifying the equilibrium conditions that lead to their formation. The needed information is not merely the existence of sub-critical entities, since this is considered a matter of fact, but their size distribution and nature (e.g., density) before the critical point, which can reflect the existence of a preferential size or a statistical distribution of size or even a different scenario.

Synchrotron-based scattering techniques play an important role in extracting information from entities suspended in the liquid phase. They are very powerful techniques thanks to the high brightness and flux and modern photon counting detectors. X-ray scattering is sensitive to the sample's electron density spatial variations. In particular, SAXS is sensitive to density variations on the scale of nanometers to micrometers, providing very strong, meaningful, and informative scattering information at small angle,[18] yielding information on particles size and their size distribution.[19,20] At higher angles (WAXS) the scattering yields information on a subnanometric scale, such as the atomic structure, more detailed information on crystallinity and defects, and more detailed size and shape distribution information. A limiting factor of WAXS is that – typically – the diffraction signal is weaker by several orders of magnitude with respect to SAXS, which therefore remains the only method exploitable when a combination of weak contrast and tiny concentrations characterizes the sample, as is the case for near-saturation $CaCO_3$ solutions. The scattering process follows the reciprocity law: scattering information from larger features corresponds to lower angles.[21,22] Therefore, the hardware used for such measurements needs to be carefully designed in order to define the optimal trade-off among energy and flux of the beam, geometry, and detector resolution.

Ballauff et al. demonstrated the effectiveness of SAXS specifically on the investigation of ACC, applying a stopped-flow device.[19,20] With that setup, two solutions of calcium and carbonate ions, both in the concentration range between 7 and 9 mM, were rapidly mixed in equal volumetric proportion and transferred into a quartz capillary. The formation of ACC – and its transformation into crystalline phases – was investigated over time at $pH = 10.5$ and $T = 25$ °C. Analyzing their experimental conditions, considering the most recent thermodynamic database [23] and the value for ACC solubility in the described condition ($\approx 1.80 \times 10^{-8}$),[15] the experimental saturation levels corresponds to values of 57 and 74, respectively (Supporting Information). It is worth noting that the saturation is here calculated as $S = IAP/Ksp$ where IAP stands for Ionic Activity Product and $Ksp$ is the solubility product of ACC. The same experimental conditions were used in the work of Huber et al., where the ACC formation was investigated by static light scattering in a cuvette.[13,14]

At these very high saturation levels, the spinodal decomposition occurs, the ACC formation rate is very high, the solution mixing plays a fundamental role on the obtained particle size distribution, the ACC transforms to crystalline phases in a time scale of tens to hundreds of seconds, and the heterogeneous nucleation on the reactor wall is highly probable, be it a capillary or a cuvette. Consistently, experimental evidences associated with the aforementioned processes were addressed by the authors.[13,14,19,20] Such experimental conditions can be effectively utilized when the aim of the study is focused on the ACC particles, their size and nature, as well as their dynamic transformation into crystalline phases. On the contrary, in the present work, we focus our study on a different timeframe, i.e. that corresponding to the stage *before* the critical condition for ACC particles formation by homogeneous nucleation. Therefore, the saturation level needs to be much closer to the unity and the system studied in static (and not dynamic) mode. The investigation of the system at saturation levels down to values < 4 (or even < 1, i.e., undersaturated) poses experimental challenges. Firstly, the concentration of the entities objective of the investigation strongly decreases (up to two orders of magnitude lower with respect to the work of Ballauff et al.). In order to achieve an appropriate signal-to-noise ratio, long data acquisition time (several minutes) at constant conditions of *S, pH* and *T* need to be accomplished. Therefore, any stopped-flow techniques are precluded, being designed to investigate dynamic processes. Secondly, due to the labile signal of the amorphous entities target of the investigation, the background generated by the sample interrogated by the beam needs to be minimized, the influence of the beam on the sample excluded (beam damage), and heterogeneous nucleation on wall of the capillary (or any other cell) ruled out.

To face this experimental challenge, a pulsation-free micrometric-size reactive horizontal liquid-jet setup was built. The liquid jet is not confined by a solid, thus it is a contactless setup not susceptible to wall effects,[24–27] the liquid interrogated by the beam is continuously renovated and beam damage is intrinsically excluded, and the dynamic mixing conditions allow a time-independent setting of concentrations of chemicals, *T, pH, S*, and ionic strength. Moreover, an accurate chemical speciation model is mandatory for a precise calculation of the system saturation level. We applied the developed thermodynamic model for ACC precipitation, which can solve the speciation as well as compute the solubility of ACC, in the specific experimental conditions.[15] Besides, the setup was calibrated for absolute scale with a standardized nanoparticle suspension.[28]

In this paper we present both the results on cluster size distribution as a function of the saturation level with respect to ACC and the appropriate setup to carry out such in-situ measurements also on other systems.

## 2. MATERIAL AND METHODS

Calcium chloride, sodium hydroxide, and sodium bicarbonate were purchased from Sigma Aldrich (analytical grade, ReagentPlus). Aqueous solutions $CaCl_2$ ($2 \times 10^{-3}$ mol/L), NaOH ($5 \times 10^{-3}$ mol/L), $NaHCO_3$ ($20 \times 10^{-3}$ mol/L) and pure water were prepared using $CO_2$-free milliQ water. A pulsation-free micrometric-size horizontal reactive liquid jet setup was specifically built for the measurements. The system as a whole was composed of four HPLC pumps – each of them equipped with a pulsation damper system and high precision Coriolis liquid mass flowmeters, a mixing system, a delay loop, and a catcher (Figure 1). A micromixer manifold equipped with five inputs and one output was used to mix the solutions. The fifth input was connected with an additional HPLC pump delivering a 10 *wt.%* acetic acid solution, which serves to clean the system. The manifold exit holds the delay loop, which consists of a Teflon tube of a certain length and internal diameter. The loop defines the delay time between the mixing point and the irradiated liquid jet. Delivering tubes before the mixer and the delay loop were thermostated using a double-walled water-jacketed tubing system. The delay loop outflow was connected to a second manifold and a capillary (nozzle), which can be chosen between different materials and internal diameters. The delay time was fixed to 75 s. The internal capillary diameter defines the overall flow rate for a defined pressure drop. In this study, stainless steel capillaries with 250 µm internal diameter and an overall flow rate of 8 mL min$^{-1}$ were used. The pumping system was remotely controlled and monitored. A catcher, collecting the ejected liquid after X-ray exposure, was equipped with a micro stirrer, a *pH* electrode, and a PT1000 sensor.

Experiments were carried out at different flow rates for each delivery line in order to achieve the specific *pH* and *S* values. The flow rates were preliminarily calculated based on the thermodynamic model and verified through the experimental *pH* values within an accuracy of ±0.01 *pH* unit. The speciation model considers the formation of $CaCO_3^0$ ion pairs, which may form clusters, and it excludes the formation of the ACC phase.[15] The agreement between the predicted and the experimental *pH* values confirm that ACC precipitation does not occur. In fact, in the case of ACC formation, the *pH* decreases rapidly and a higher amount of NaOH would be needed to keep the *pH* at the set point. Consistently, ACC precipitation occurs if *S* is increased above the critical point, readily recognized by the *pH* probe in the catcher. These experimental evidences confirm the validity of the thermodynamic model and exclude the formation of ACC in the experimental conditions. We conclude that to guarantee that the *S* values are correctly set, an accurate thermodynamic model of the system, an accurate calibration of the *pH* meter (calibrated close to the *pH* of interest and at the same *T* and ionic strength of the experiment), and the precise concentrations of chemicals are needed. Moreover, the precise on-line monitoring of *pH* and *T* and an accurate pulsation-free flow rate control of the four chemicals are fundamental to keep the saturation of the system constant during the entire measurement timeframe.

Table 1 summarizes the experimental conditions and the resulting saturation levels. Four experimental conditions were analyzed in details hereafter named as $C_0$, $C_2$, $C_4$, and $C_8$ for $0 \leq S \leq 3.02$. These values define experimental conditions before the critical point for ACC homogeneous nucleation, which corresponds to $\approx 3.7$ at $pH=9.00$ and $T=25.0$ °C (Figure 1D, in ref. 15). In particular, $C_2$ corresponds to undersaturated conditions and $C_0$ represents the blank, periodically measured as background signal.

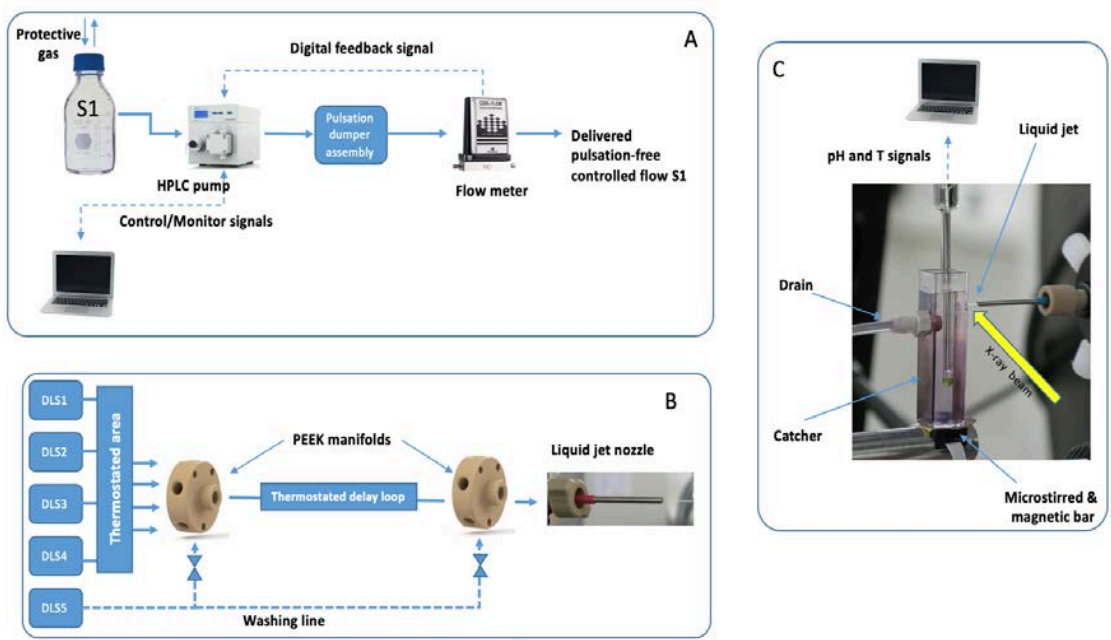

Figure 1: Pulsation-free micrometric-size horizontal reactive liquid jet setup mounted at the materials science beamline (X04SA-MS) for the in-situ X-ray scattering experiments. (A) Delivery lines for the solution S1 (DLS1). The chemical S1 is stored under atmosphere protected glass bottle and delivered by the HPLC pump to a pulsation damper device and then to a Coriolis mass flow meter. The latter generates the feedback signal to the delivery pump. As a result, a remotely controlled liquid flow up to 10 mL min$^{-1}$ with an absolute error of ±0.01 mL min$^{-1}$ is achieved. (B) Five delivery lines (DLS1 to DLS5) are implemented in the setup; one of them serves as washing line (DLS5), whereas the other four deliver the chemical as mentioned in the main text. The chemicals are mixed in a first manifold device. Then, a delay loop of 75 s is applied. A second manifold device hosts the nozzle for the liquid jet. (C) Detail of the generated liquid jet through a nozzle (250 µm internal diameter) where the total flow rate is set to 8 mL min$^{-1}$. The liquid, magnetically stirred, is collected in a catcher that hosts a micro *pH* electrode and a Pt1000 sensor and it is continuously drained. The X-ray beam interrogates the liquid jet as indicated by the yellow arrow. The SAXS signal is collected by a Mythen II detector. The measurement in progress in the figure refers to the calibration procedure with diluted colloid of Au nanoparticles (slightly pink solution in the catcher).

Table 1 – Calculated saturation levels with respect to ACC solubility ($Ksp = 2.58 \times 10^{-8}$) for the four experimental conditions considered.[15] Temperature, $pH$, overall flow rate, and delay time were fixed to 25.0 °C, 9.00, 8.00 mL min$^{-1}$, and 75 s, respectively. The concentration of the chemicals CaCl$_2$, NaOH, and NaHCO$_3$ were 2.000, 4.984, 20.014 × 10$^{-3}$ mol/L, respectively.

|  | **C$_0$** | **C$_2$** | **C$_4$** | **C$_8$** |
|---|---|---|---|---|
| CaCl$_2$ flow rate (mL/min) | 0.00 | 0.40 | 0.80 | 1.61 |
| NaHCO$_3$ flow rate (mL/min) | 3.72 | 3.67 | 3.62 | 3.52 |
| Water flow rate (mL/min) | 0.90 | 0.92 | 0.95 | 1.00 |
| NaOH flow rate (mL/min) | 3.38 | 3.00 | 2.63 | 1.88 |
| Calcium total (mol/L) | 0.00 | $1.00 \times 10^{-4}$ | $2.01 \times 10^{-4}$ | $4.02 \times 10^{-4}$ |
| Carbon total (mol/L) | $9.31 \times 10^{-3}$ | $7.35 \times 10^{-2}$ | $7.24 \times 10^{-2}$ | $7.04 \times 10^{-2}$ |
| Ca/C ratio (-) | 0.00 | $1.09 \times 10^{-2}$ | $2.22 \times 10^{-2}$ | $4.56 \times 10^{-2}$ |
| Calcium ion (mol/L) | 0.00 | $7.97 \times 10^{-5}$ | $1.61 \times 10^{-4}$ | $3.26 \times 10^{-4}$ |
| Carbonate ion (mol/L) | $5.60 \times 10^{-8}$ | $5.60 \times 10^{-4}$ | $5.60 \times 10^{-4}$ | $5.59 \times 10^{-4}$ |
| Act. coef. calcium ion | - | $6.65 \times 10^{-1}$ | $6.63 \times 10^{-1}$ | $6.60 \times 10^{-1}$ |
| Act. coef. carbonate ion | $6.54 \times 10^{-1}$ | $6.53 \times 10^{-1}$ | $6.51 \times 10^{-1}$ | $6.48 \times 10^{-1}$ |
| Calcium ion activity (mol/L) | - | $5.30 \times 10^{-5}$ | $1.07 \times 10^{-4}$ | $2.15 \times 10^{-4}$ |
| Carbonate ion activity (mol/L) | $3.66 \times 10^{-8}$ | $3.66 \times 10^{-4}$ | $3.65 \times 10^{-4}$ | $3.62 \times 10^{-4}$ |
| CaCO$_3$ ion pair (ppm) | 0.00 | 2.04 | 3.96 | 7.44 |
| IAP (mol/L)$^2$ | - | $1.94 \times 10^{-8}$ | $3.88 \times 10^{-8}$ | $7.80 \times 10^{-8}$ |
| Saturation (-) | **0.00** | **0.75** | **1.51** | **3.02** |

Synchrotron Small-Angle X-ray Scattering (SAXS) measurements were carried out at the Material Science beamline (X04S-MS) of the Swiss Light Source (SLS) at PSI.[29] This synchrotron station is built for WAXS powder diffraction measurements but it also has some SAXS capabilities. Modern photon counting detectors (Mythen II & PILATUS 6M) are installed. The liquid jet, horizontal and orthogonal to the X-ray beam, was mounted on a double micrometric translation stage and optically centered with respect to the diffractometer circle by a high-resolution camera (Figure 1C). The X-ray beam was set at 9.5 KeV ($\lambda$=1.305 Å) where the X-ray flux is maximal.

Data were collected with the Mython II detector system,[30] that with its 0.0036 deg pixel, has a sufficient resolution for SAXS on this system with a minimum accessible $2\theta$ scattering angle of ~ 0.18 - 0.20 deg, corresponding to a minimum accessible momentum transfer of $Q=4\pi \sin(\theta)/\lambda=0.016$ Å$^{-1}$ where $\lambda$ is the wavelength of incident X-ray.

## 3. DATA ANALYSIS

Before their analysis, the experimental data need a careful conversion, from raw data (photon counting) into diffraction data (intensity-angle; intensity-momentum transfer Q on regular step). Scattering patterns were collected with reasonable acquisition times from ACC solutions at the selected saturation levels, as well as from pure water, and from air. The system was washed with acetic acid, followed by water, before each acquisition. Ancillary data (S, time-delay between the mixing point and measuring point, T, pH, and jet size) were also recorded. In following sections, the applied calibration on the SAXS data and fitting approaches are described. As most usual, the scattering data were collected in arbitrary units. A suitable standard was measured in order to obtain an appropriate scale calibration factor and convert the data to absolute units. This standard was a spherical gold nanoparticle (AuNP's) suspension of known concentration (200 mg/l as Au) and narrow size distribution (18 nm, evaluated by TEM),[31] measured using the same liquid jet and instrument setup as for the samples under investigations. The scale calibration factor was evaluated to $1.011 \times 10^{-17}$.[28]

### 3.1.1 FITTING APPROACHES

The entities in suspensions were assumed to consist of polydisperse particles with a spherical shape.[32] A parametric distribution model was used for modeling and analyzing the collected data.[33]

Two approaches were attempted. Firstly, a more classical approach was evaluated. We assumed a log-normal size distribution of spherical clusters of unknown density. The scattering intensity of this polydisperse system is weighted by a two-parameter log-normal distribution. The observed data (with a measured background pattern whose relative scale was set free) for the saturation levels ($C_2$, $C_4$) were fitted very nicely with the model intensity, but the results presented several weak points (extremely broad distributions, little correlation between samples, not all fits were converging). This approach was rejected in favor of a slightly more conceptually complex one (bimodal approach), albeit without increasing the number of parameters.

### 3.1.2 BIMODAL APPROACH

In this approach, we assumed that the whole system is composed of two populations of particles: one population includes monodisperse small dense clusters (SDC hereafter), with diameter of about 2 nm; a second population includes polydisperse aggregates (i.e., superclusters) with higher volume, formed by several SDC loosely packed and without coalescence. Reasonable values of packing fraction range between 50% and 72%. The lower limit still gives some compactness; the upper limit is the close-packing limit, which is still below the coalescence region. We assume a packing fraction of 70% (vol.) of SDC in the aggregates, near the close-packing limit, and a minimum diameter of ~ 5 nm (i.e., a supercluster contains at least 8 SDC). The contributions of these two populations lead to two linear components for the intensity (one from the population of the SDC and one from the population of the superclusters), plus the separately measured background ($C_0$).

The term due to the SDC, as they result to be rather small, is a relatively flat and featureless trace in the experimental $Q$ range (0.015-0.07 Å$^{-1}$) and it is weak and highly correlated with the background. Therefore, (i) the scale of this trace cannot be precisely determined and (ii) the SDC diameter – approximated as monodisperse spheres – can be determined within a 20-30% error, which may also be intrinsic in the SDC nature. Nevertheless, the fact that this signal is refined to consistent values (diameter of about 2 nm, corresponding broadly to the monomers detected by Huber et al.) in all of our best experimental patterns (8 series of data) for all experimental conditions, adds credibility to their existence, at least qualitatively.

For the supercluster (polydisperse aggregates), the size distribution function has been refined both as a lognormal and an exponential distribution (which is a limiting case of the Schulz-Flory a.k.a. Gamma distribution [34–36]):

$$P_m = C \exp(-D_m/D_0), \qquad (1)$$

the latter distribution being then selected as giving the most credible results. Here $C$ is a normalization constant such that the number fractions $P_m$ sum to 1, $D_0$ is the distribution parameter, $D_m = m\delta$ is the diameter of the $m^{th}$ cluster (variable between 5 and 100 nm) and $\delta$ is a convenient step representing the diameter of a sphere containing one SDC and an equal volume of water. Here, the role of $D_m$ is more a computational detail than a meaningful physico-chemical detail. Hereby, we are not making a strict structural hypothesis on the superclusters. In fact, when dealing with continuous size (diameter) distributions of particles, one must evaluate the distribution at discrete values of the diameter $D$, which are finely enough spaced to give meaningful results but not so much as to overburden the calculations. In our case of larger spheres (i.e., spheres built as loose coalescence of much smaller predefined spheres), such diameter steps are chosen so that in the $m^{th}$ spherical shell of thickness $\delta/2$ there are $3m^2 - 3m + 1$ smaller particles, and accordingly, in the $m^{th}$ sphere of diameter $D_m = m\delta$ there are $m^3$ smaller particles. This treatment represents a robust and convenient way of defining the sampling step for size distributions.[37] Accordingly, to evaluate $\delta$, we use a value (70%) of the packing density that is within the most meaningful region (30% to 72% for loose to close packing without coalescence).

The scattering intensities must be summed over all the particles sizes and weighted by their size distribution function $P_m$. Hence, the calculated scattering intensity of this polydisperse system is given by

$$I(Q) = k\, N\, (\Delta\rho)^2 \sum_m P_m\, V_m^2 \left(3\, \frac{\sin(QR_m) - QR_m \cos(QR_m)}{(QR_m)^3}\right)^2 \qquad (2)$$

where $m$ is indexing over the possible different spherical clusters (see below), $N$ is the total number of particles in the beam, $R_m = D_m/2$ is the sphere radius, $Q$ is the transferred momentum, $V_m$ is the volume of the clusters, and $\Delta\rho$ is the clusters electron density contrast. At $Q = 0$, the scattering forward intensity is given as

$$I(Q=0) = k\, N\, (\Delta\rho)^2 \sum_m P_m\, V_m^2 \qquad (3)$$

with $k$ representing the calculated scale factor to bring absolute intensities onto the experiment scale, estimated by a reference material as mentioned above. In order to fit the data with the model, the free parameters were the exponential size distribution ($D_0$), the scale factor of the large aggregates distribution ($S_{agg}$), the scale of the background trace to be subtracted ($S_{bkg}$), and the scale of an additional diffraction trace of a fixed small-diameter sphere representing the SDC ($S_{SDC}$). The scales $S_{bkg}$ and $S_{SDC}$ always resulted highly correlated (within $10^{-4}$ from perfect correlation), so their separate values have no meaning. As such, the scale $S_{SDC}$ does not allow quantification of free-standing SDC. However, eliminating their contribution from the fit has negative effect on the stability of results, therefore they were kept in, at least as an additional background term; it is likely that their contribution is present, but they cannot be quantified. The scale of the measured background always resulted very close to 1 as expected. Only small and unaccounted for parasitic effects (e.g., tiny jet diameter variations, monitor error) could affect it. A simple grid search algorithm dealt with finding optimal values of the distribution parameter, while scale factors were separately optimized at each step as a linear minimization problem. The goodness of fit and the other relevant factors were also evaluated. Figure 2 shows the best fit obtained for the data of the selected conditions ($C_2$, $C_4$, $C_8$).

As indicated in eq. 3, the scattering intensity is proportional to the square of the scattering contrast, which represents the excess electron density of the suspended particles with respect to the medium surrounding medium. SDCs are poly-hydrated $CaCO_3$ ion pair; the composition is generally $CaCO_3 \cdot nH_2O$ with atomic weight of $100.1+18.02n$ and $50+10n$ electrons per unit formula.

Water was found to be a key factor for ACC nucleation.[38] This raised a suggestion of using a different water fraction for every saturation level. The water content and the cluster density at each condition ($C_n$) were estimated applying the thermodynamic model [15] which take into account for different water content. We fixed 22 $CaCO_3$ formula unit per cluster [39] and a number of water molecules per Ca atom of 18, 12, 8 for conditions $C_2$, $C_4$, and $C_8$, respectively. The calculated SDC mass densities for conditions $C_2$, $C_4$, and $C_8$ are estimated to be 1.160, 1.227, and 1.314 g/cm$^3$, corresponding to SDC sizes as 2.94, 2.62, 2.35 nm, respectively (Supporting Information). Therefore, SDCs have a distinct $\Delta\rho$ at each condition, and the contrast of the respective superclusters is $\eta\Delta\rho$ where $\eta$ is the packing fraction (which we arbitrarily fixed to 0.7 – near the close-packing fraction). Note that also the packing fraction could also very well be variable with the saturation. However, this at the moment must remain speculative.

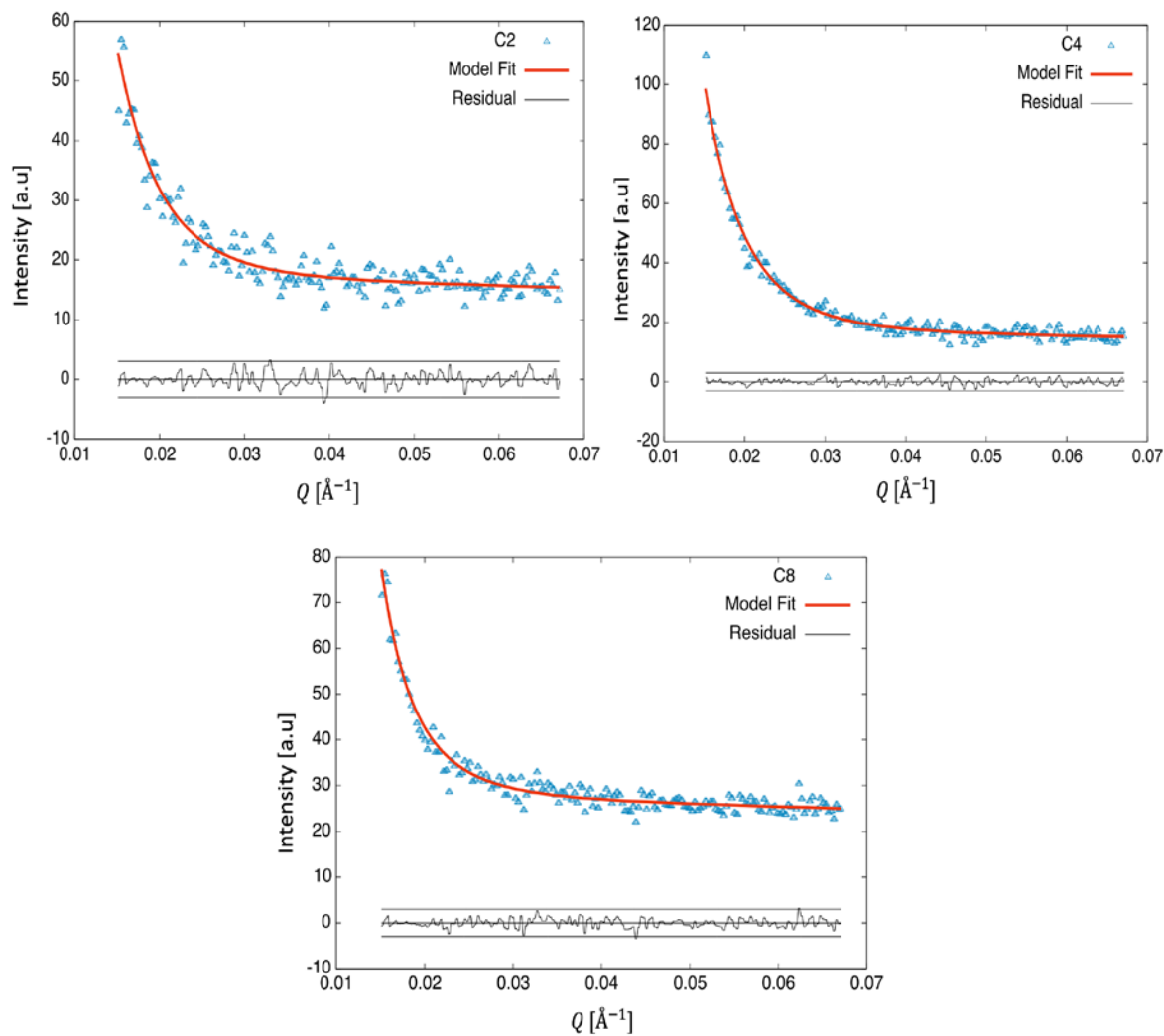

Figure 2: Best fit obtained for conditions $C_2$, $C_4$, and $C_8$ (GOF: 1.15, 0.97, and 0.99 respectively) using the bimodal approach (SDC and aggregates with an exponential size distribution). Insets: residual divided by the standard deviation σ, bracketed by the ±3 σ lines.

Although our data are noisier than usual for a SAXS experiment (because of the very low concentrations and contrast), they showed to be informative and selective enough to clearly show a preference for an exponential distribution of superclusters with respect to a single-size and even with respect to a classical log-normal. The exponential distribution fits result quite nice from any perspective, meaningfulness of results, graphical look, and especially the statistical indicators. In this respect, including the contribution due to SDC as aforementioned discussed, the goodness-of-fit, GoF, is close to 1 (the ideal value, see table 1) in all cases; moreover, practically no point deviates more than $3\sigma$ from the refined model value, as shown in Figure 2.

## 4. RESULTS AND DISCUSSION

The reliability and significance of the experimental SAXS data were checked for all the measurements conducted at the different saturation conditions using the two approaches discussed above. The subject of our study is low saturation (S<4) calcium carbonate solutions. In our experimental condition, the IAP is $< 8 \times 10^{-8}$ and the estimated concentration of $CaCO_3^0$ ion pair is < 8 ppm (table 1).[15] Therefore, assuming that all ion pairs form SDC and considering the lower scattering contrast of entities that might be formed in these conditions, a meaningful signal must be 2-3 orders of magnitude weaker relative to the liquid carrier signal. With respect to Ballauff et al. and Huber et al., our concentration range is set more diluted up to two orders of magnitude and the cluster electron density contrast was also about 2-3 times smaller. While we still obtain a useful signal, with robust signal-to-noise ratio in the Guinier region and in the Fourier region, the Porod region signal is too low to apply the Porod invariant method.[40,41]

We also attempted a quantification of $Ca^{2+}$ contained in the superclusters. The SDC were ignored, as not quantifiable. For the superclusters, the scale and the exponential diameter distribution parameter could be properly refined. As a summary, we could evaluate:

- the concentration of Ca ions ($C_{Ca^{2+}}$) in superclusters to be compared with the concentration of $CaCO_3^0$ ion pair present in solution ($C_{CaCO3}$);
- the average diameter weighted by the mass distribution $< D >_{MD}$;
- the average width of the mass distribution (r.m.s. width-MD);

- the diameter at maximum of the mass distribution;

- the average diameter weighted by the number distribution $<D>_{ND}$;

- the average width of the number distribution (r.m.s. width-ND);

- the exponential distribution parameter ($D_0$);

- the scale factors of the SDCs ($S_{SDC}$), superclusters ($S_{aggr.}$) and background ($S_{bkg}$).

Moreover the correlations of:

- the background with superclusters ($I_{bkg-aggr}$);

- the background with SDC ($I_{bkg-SDC}$);

- the aggregates with SDC ($I_{SDC-aggr}$);

and

- $I_{aggr}(Q=0)$, $I_{SDC}(Q=0)$, and the product $Q_{min} \times R_g$.

Most of these values are reported in Table 2 for the selected conditions $C_2$, $C_4$, and $C_8$.

**Table 2:** Summary of the fit results obtained (fixing arbitrarily a packing fraction is 0.7) from the bimodal approach for the selected supersaturation levels ($C_2$, $C_4$, $C_8$). The calculated parameters are defined in the text. ND, MD indicate the number distribution or the mass distribution, respectively. The intensities extrapolated at $Q=0$ for the SDC and aggregates contributions are given in lieu of the raw scale factors. $R_g^*$ is simply the gyration radius calculated for a sphere having a diameter equal to $D_0$ – the distribution parameter – as it is useful to evaluate the data extension in the Guinier region (where $Q R_g^* < 1.5$).

| Parameter ($\eta = 0.7$) | $C_2$ (S=0.75) | $C_4$ (S=1.50) | $C_8$ (S=3.04) |
|---|---|---|---|
| $CaCO_3$ in aggr. [ppm] | 1.276 | 3.787 | 1.506 |
| $CaCO_3$ % in aggr. | 62.6 | 95.6 | 20.2 |
| GoF | 1.16 | 0.95 | 0.99 |
| Parameter of the Exp. Distribution ($D_0$) [nm] | 7.35 | 10.63 | 15.60 |
| $<D>_{ND}$ [nm] | 12.44 | 15.13 | 20.00 |
| r.m.s width-ND [nm] | 7.29 | 10.60 | 15.98 |
| Diameter at maximum MD [nm] | 22.32 | 32.18 | 48.26 |
| $<D>_{MD}$ [nm] | 29.51 | 42.56 | 63.99 |
| r.m.s width-MD [nm] | 14.63 | 21.24 | 31.98 |
| $S_{bkg}$ | 0.995 | 0.981 | 0.992 |
| Correlation ($I_{bkg-SDC}$) | -0.9999 | -0.9999 | -0.9999 |
| $I_{aggr}(Q=0)$ | 342.2 | 4052.0 | 7502.0 |
| $I_{SDC}(Q=0)$ | 16.47 | 63.08 | 26.75 |
| $R_g^*$ [nm] | 2.84 | 4.11 | 6.19 |
| $Q_{min}$ x $R_g^*$ | 0.43 | 0.62 | 0.93 |

As discussed, the experimental evidences reveal that the SDC signal is poorly defined, being it a rather flat trace highly correlated with the background. Therefore, size and quantity of SDC are affected by too high incertitude to be considered meaningful. On the other hand, we can quantify Ca ions from the superclusters because the signal is much more structured (*i.e.*, the large entities can be easily and precisely quantified). Consequently, the Ca ions concentration ($C_{Ca}$, Table 2) – omitting the unknowable fraction of the SDC – is below the calculated value ($C_2$), very near and slightly above it ($C_4$) and very underestimated ($C_8$, probably because of the formation of large entities). In this sense, we assumed a unimodal population of small ($\approx$2nm) (with density 1.16 to 1.31 g cm$^{-3}$, Supporting Information) spherical clusters and a population of less dense but larger superclusters (broad size distribution up to 100-200 nm) constituted by the SDC with approximately 30% water existing in between them. A reasonable packing fraction limit of about 70% was assumed, although a larger and possibly variable packing fraction would be more consistent (for $C_2$ and $C_4$ at least) with the estimated $CaCO_3^0$ ion pair concentration by the thermodynamic model. The presence of a population of SDC seems consistent with other experimental results [11,17,42] and compatible with the calculated critical size. [15] On the other hand, the presence of highly hydrated large superclusters seems consistent with a liquid-like separation model.[39] Therefore, within the approximation and assumption included in the data analysis, combined with the intrinsic limitation of technique and the properties of the material under investigation, we can state that the experimental results are consistent with the presence of clusters characterized by a diameter of about 2 nm that might be in a high numerical concentration but almost negligible overall mass, and a population of superclusters, limited in number concentration but containing the majority of the $CaCO_3^0$ ion pair in solution.

For a so described system, at the early stage of the denser matter formation, it is not a surprise that by cryoTEM the limited number of large superclusters might be excluded by sampling and they are not detectable by AUC. On the other hand, the computational model might emphasize the presence of superclusters (or nanodroplets of denser liquid phase), corresponding to the form in which the majority of $CaCO_3^0$ ion pair mass is accumulated.

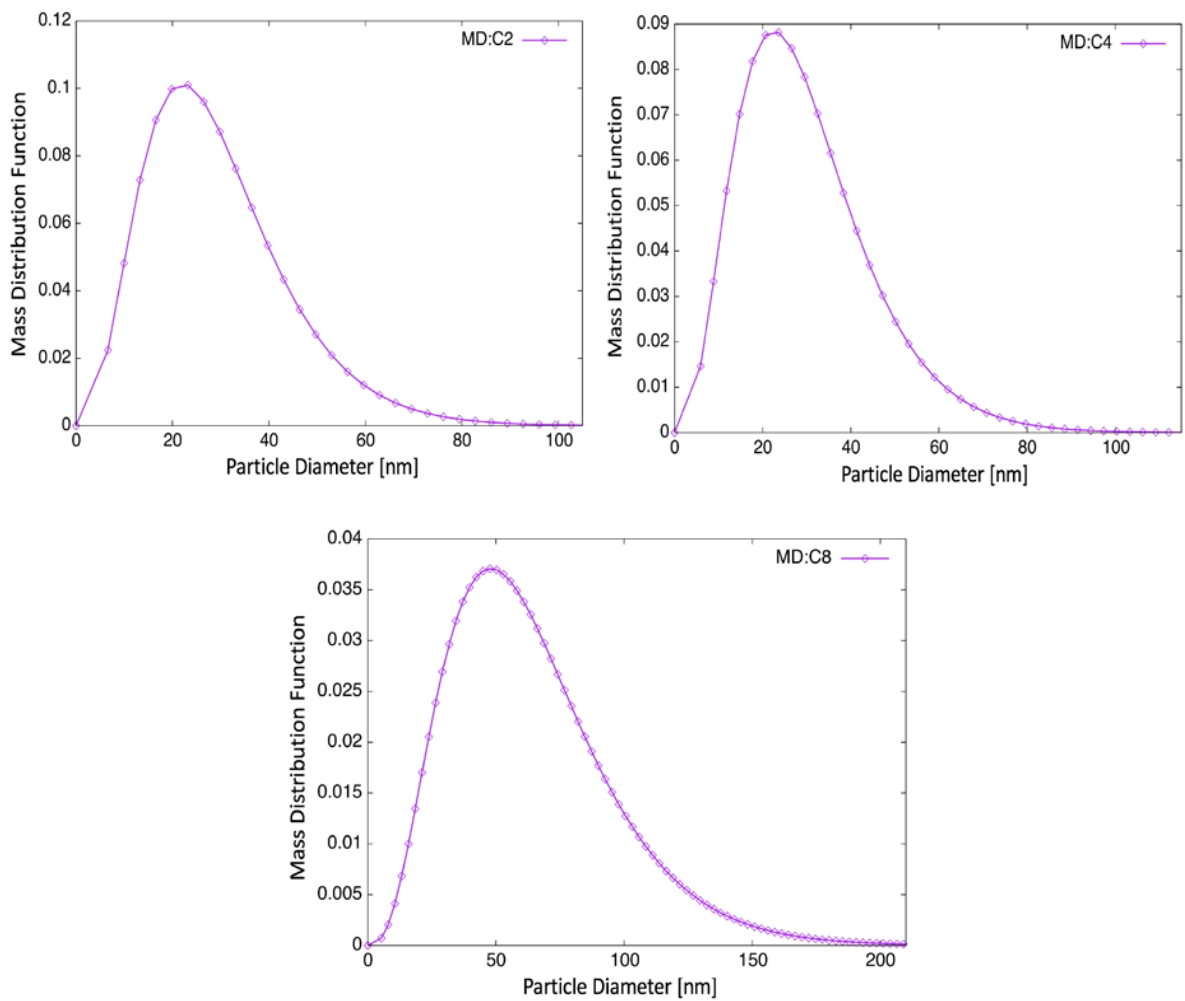

**Figure 3.** The obtained mass distribution of the objects ($C_2$, $C_4$, $C_8$). The distribution functions were obtained based on the bimodal approach.

The beauty of the presented techniques and the data analysis approach is that the aqueous system under observation is not perturbed by any sampling or manipulation and the direct evaluation of superclusters at constant $S$, $pH$, and $T$ were achieved. Moreover, even if affected by large incertitude, the presence of SDC is also consistent with the experimental results. The same approach can be extended and applied to other systems, paving the way toward a more comprehensive understanding of the early stage of solid formation and merging only apparently divergent conclusions based on other investigation (and complementary) techniques.

In the specific case of calcium carbonate formation pathway, before ACC nucleation, we speculate an increase of supercluster density with the saturation level and therefore the involvement of water in the formation pathway. In the literature, Raiteri et al. have already reported the influence of water as key factor for the nonclassical nucleation of ACC.[38] The authors concluded that their calculations are consistent with the formation of a narrow distribution of small clusters during the early stage of the process, that then agglomerate to larger entities. A Gibbs free energy landscape giving more than one point with zero derivative, so implying a nonclassical nucleation pathway, is also proposed.

Our experimental data are consistent with the presence of small and large clusters, but the representation of the process is formally different. Figure 4 summarizes the overall concept for the ACC formation pathway from the point of view of the solution chemistry and population balance approach. [15] The black line represents the measured free $Ca^{2+}$ ion in solution, the violet line is the overall amount of $CaCO_3^0$ ion pair (as isolated entity and clusters) and the blue line reports on the amount of ACC formed. The red circles represent the experimental conditions discussed in this paper, all belong to the stage I of the solid formation process, i.e. under (pseudo)thermodynamic equilibrium and before ACC nucleation. The dashed line denotes the solubility limit of ACC, experimentally evaluated from the plateau of the free $Ca^{2+}$ curve for time> 120 min. Indeed, C2 corresponds to undersaturated conditions (below the dashed line) whereas C4 and C8 correspond to supersaturated conditions, but with S lower than the critical point for ACC nucleation (maximum of the black curve). In the figure, the reported squared areas depict the populations of SDCs and superclusters in solution at each stage. Even when the system is undersaturated, both SDCs and superclusters exist. The system is highly hydrated, the SDC themselves are denser than water and aggregated in hydrated superclusters with a broad size distribution. As the system saturation increase, both the density of the SDC and the size of the superclusters increase. This process gradually proceeds until the critical conditions for primary nucleation, which corresponds to a critical density of the SDC within the superclusters. In the figure, the yellow particle denotes the primary nucleation event, i.e. the transformation of a SDC to an ACC primary nucleus by dehydration. Almost simultaneously, secondary nucleation occurs within the superclusters (red spheres), leading to the massive ACC formation. The process after the maximum of the black curve is described in details in ref 15. Here, we complete that investigation with the conditions $C_2$, $C_4$, and $C_8$.

Even if we are not able to figure out the details of the Gibbs free energy landscape, it seems clear that when the saturation of a system is gradually raised, the system reacts defining new (pseudo)equilibrium states at each *S* levels, where a population of entities with a large variety of size and density is present. With this picture in mind, the representation of the Gibbs free energy landscape, as a well-predefined and fixed profile that is carefully followed during the entire precipitation pathway, results obsolete. In fact, to each *S* value, a new function for both the volume and the surface contributions to the overall Gibbs free energy –and therefore new size distribution of denser entities– corresponds. Thus, the appropriate Gibbs free energy landscape cannot be plotted in a 2D plane, since it is function of size and saturation, outlining a more complex –at least 3D– representation for the solid formation pathway.

It is worth mentioning that, according to schematization of Figure 4, the equations associated to the classical nucleation theory can still be used to estimate the nucleation rate since they are unaffected by the details of under-critical events.

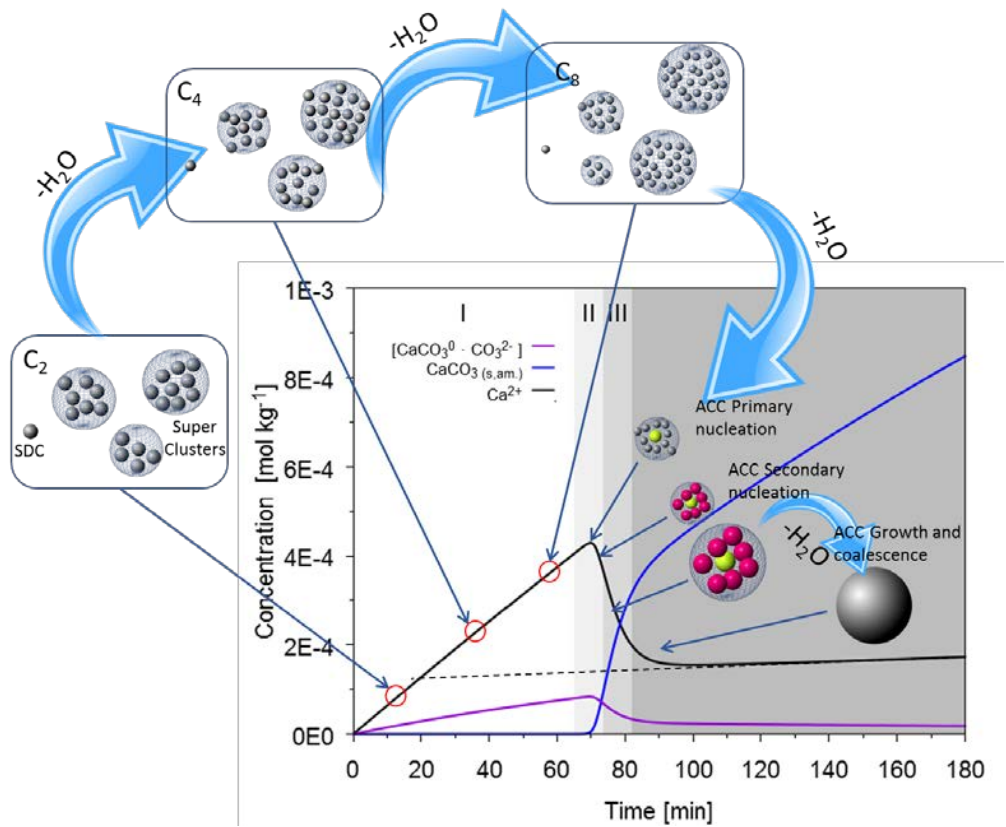

Figure 4: Comprehensive solid formation pathway based on the experimental evidences collected in this work, combined with previous finding and literature consistency.[15,38]

5. CONCLUSION

The SAXS technique combined with a liquid jet setup, an accurate thermodynamic model, and the bimodal approach have allowed the detection of superclusters in-situ. As a model system, the calcium carbonate formation pathway was investigated at different saturation levels, including undersaturated conditions.

The experimental results are consistent with the presence of a broad size distribution of entities (superclusters) with variable density and highly hydrated, which were postulated by the computational model. Simultaneously, a population of clusters of about 2 nm, were qualitatively detected, previously identified in the literature using cryoTEM and AUC, which are mainly aggregate in the superclusters. The presented results are not only consistent with the literature, but for the first time, both populations of denser matters in solution are simultaneously detected by SAXS, before the critical condition for ACC precipitation. Such entities are present even in undersaturated conditions (with respect to ACC). The populations of SDC and supercluster evolve increasing the saturation level towards higher density (dehydration) and larger superclusters.

Because size and composition of superclusters change with the saturation, the conceptual representation of the Gibbs free energy landscape should be revisited according to a surface in a three-dimensional space. Therefore, the concept of CNT and NCNT should be entirely reconsidered.

ASSOCIATED CONTENT

Supporting Information: 1. Estimation of SDC size and hydration state; 2. Estimation of the saturation level of refs 19-20.

AUTHOR INFORMATION

* Andrea Testino. Paul Scherrer Institut, Energy and Environment Research Division, Villigen PSI, CH-5232 Switzerland. Email: andrea.testino@psi.ch

AUTHOR CONTRIBUTIONS

The manuscript was written through contributions of all authors. All authors have given approval to the final version of the manuscript. ‡These authors contributed equally.

NOTES

The authors declare no competing financial interest.